\documentclass[preprint,amsmath,amssymb,aip,cha]{revtex4-1}
\usepackage{amssymb}
\usepackage{graphicx}
\usepackage{epstopdf}
\usepackage[bf,SL,BF]{subfigure}
\usepackage{CJK}

\begin{document}
\title{Energy thresholds of discrete breathers in thermal equilibrium and relaxation processes}
\author{Yi Ming}
\email{meanyee@mail.ustc.edu.cn}
\author{Dong-Bo Ling}
\affiliation{School of Physics and Material Science, Anhui University, Hefei, Anhui 230601, People's Republic of China}
\author{Hui-Min Li}
\affiliation{Supercomputing Center, University of Science and Technology of China, Hefei, Anhui 230026, People's Republic of China}
\author{Ze-Jun Ding}
\email{zjding@ustc.edu.cn}
\affiliation{Department of Physics and Hefei National Laboratory for Physical Sciences at the Microscale, University of Science and Technology of China, Hefei, Anhui 230026, People's Republic of China}%
\affiliation{Key Laboratory of Strongly-coupled Quantum Matter Physics, University of Science and Technology of China, Chinese Academy of Sciences, Hefei, Anhui 230026, People's Republic of China}
\date{\today}

\begin{abstract}
So far, only the energy thresholds of single discrete breathers in nonlinear Hamiltonian systems have been analytically obtained. In this work, the energy thresholds of discrete breathers in thermal equilibrium and the energy thresholds of long-lived discrete breathers which can remain after a long time relaxation are analytically estimated for nonlinear chains. These energy thresholds are size dependent. The energy thresholds of discrete breathers in thermal equilibrium are same as the previous analytical results for single discrete breathers. The energy thresholds of long-lived discrete breathers in relaxation processes are different from the previous results for single discrete breathers but agree well with the published numerical results known to us. Because real systems are either in thermal equilibrium or in relaxation processes, the obtained results could be important for experimental detection of discrete breathers.
\end{abstract}
\maketitle
\begin{quotation}
Discrete breather is an intrinsic spatially localized nonlinear excitation which can be excited in any kind of discrete nonlinear systems. They can be excited both in thermal equilibrium and in relaxation processes. Once the long-lived discrete breathers are formed in the relaxation processes, the energy relaxations are very slow. The energy thresholds of discrete breathers in thermal equilibrium and relaxation processes can be only numerically estimated so far because of the strong interactions between breathers with other excitations, for instance, phonons. They are thus regarded as identical to the previous analytical energy thresholds of single discrete breathers in Hamiltonian systems.
In this work, the energy thresholds of discrete breathers in thermal equilibrium and relaxation processes and thus of the slow energy relaxations are analytically estimated. The energy thresholds of discrete breathers in thermal equilibrium are same as the previous results for single discrete breathers. However, the energy thresholds of long-lived discrete breathers in relaxation processes are different from the previous results. Remarkably, they agree well with the published numerical results known to us. Real systems are either in thermal equilibrium or in relaxation processes. Therefore, the obtained energy thresholds, especially the size-dependence of them, are expected to be important for experimental detection of discrete breathers.
\end{quotation}

\section{Introduction}\label{sec1}
Discrete breathers are time periodic and spatially localized (typically exponentially) nonlinear excitations in nonlinear systems \cite{AUBRY1997201, Aubry20061, FLACH1998181, Campbell2004, Flach20081}. They are also referred to as intrinsic localized modes in condensed matter physics \cite{Sato2006137} and discrete solitons (or lattice solitons) in nonlinear optics \cite{Lederer20081}. Nonlinearity and discreteness, which are intrinsic to many natural systems, are two ingredients of discrete breathers. Therefore, discrete breathers have attracted intense interest in many areas of physics as well as chemistry and biology over the past three decades, see, e.g., the most recent review articles [\onlinecite{Flach20081, Sato2006137, Lederer20081}]. Intriguingly, discrete breathers have been experimentally observed in a wide variety of different media such as atomic lattices \cite{Manley2006125501, Markovich2002195301}, alkali halides \cite{Manley2009134304, Manley2011}, graphite \cite{Liang20145491}, charge-transfer solids\cite{Swanson19993288}, Josephson junction arrays\cite{Trias2000741, Binder2000745}, coupled antiferromagnetic layers\cite{Schwarz1999223, Sato2004486, Wrubel2005264101, Stock2016017201}, layered high-$T_c$ superconductors \cite{Dienst2013535}, micromechanical cantilever arrays\cite{Sato2003044102, Sato2003702, Sato2007214101}, torsionally coupled pendula\cite{Cuevas2009224101}, granular crystals \cite{Boechler2010244302}, electrical lattices\cite{Stearrett20075394, English2008066601, English2010046605, English2012084101, English2013022912}, optical waveguides and photonic crystals\cite{Eisenberg19983383, Fleischer2003023902, Fleischer2003147, Cohen2005500, Wang2006083904}, Bose-Einstein condensation\cite{Eiermann2004230401}, biopolymers \cite{Xie20005435, Barthes20025230}, and so on.

For a generic nonlinear system, it is expected that the energy of the system must higher than a critical value to make the nonlinearity become relevant. Therefore, there is an energy threshold to observe a discrete breather. Many experiments have indeed revealed the existence of the energy threshold \cite{Manley2006125501, Manley2009134304, Manley2011, Liang20145491, Trias2000741, Schwarz1999223, Sato2004486, Wrubel2005264101, Stock2016017201, Eisenberg19983383, Fleischer2003147}. The size-dependent and dimension-dependent energy thresholds of discrete breathers have been analytically obtained in nonlinear Hamiltonian systems \cite{Flach19971207, FLACH1996223, Kastner2004104301, Kastner20041923, Hajnal2008124101}. The analytical estimations can be achieved based on the fact that the studied Hamiltonian systems were supposed to be excited with nothing but one single discrete breather. Therefore, this discrete breather, which corresponds to a periodic orbit in the phase space, can be regarded as the tangent bifurcation from a phonon band edge mode \cite{FLACH1996223, Kastner2004104301, Kastner20041923}. For a very common system: a $d$-dimensional $N$-sites lattice whose Hamiltonian admits a Taylor series expansion around its equilibrium point, the tangent bifurcation can occur when the amplitude of the phonon band edge mode is higher than a critical value $\sim N^{-1/d}$. The energy threshold of discrete breather is thus $e_0\sim N^{-2/d}$ per site \cite{FLACH1996223, Flach19971207}. In the coordinate space, the amplitudes of a single breather decay in space away from the breather center. The energy threshold can be thus also estimated by linearizing the equations of motion in space away from the breather center \cite{Flach19971207} or by restricting the nonlinearity to a single bond \cite{Hajnal2008124101}.

However, many numerical results \cite{PEYRARD1998184, Flach19945018, Gershgorin2005264302, Eleftheriou200320, Ivanchenko2004120, Eleftheriou2005142} and experimental results \cite{Manley2006125501, Manley2009134304, Manley2011, Stock2016017201} indicate that discrete breathers can be observed in thermal equilibrium. Wherein the experiments Refs. [\onlinecite{Manley2006125501, Manley2009134304, Manley2011}] indicate that discrete breathers can be excited in thermal equilibrium only when the temperatures are higher than the corresponding critical values. Therefore, there also exist the energy thresholds to excite discrete breathers in thermal equilibrium. We would like to emphasise here that the energy threshold in thermal equilibrium $e_{eq}^c$ should be better regarded as the activation free energy with respect to the number of discrete breathers in thermal equilibrium as $n_{eq}\propto \exp(-e_{eq}^c/k_B T)$ \cite{Sievers1988}. In these thermalized nonlinear systems, the previous analytical methods are not suitable to estimate the energy thresholds of discrete breathers because of the strong interactions between discrete breathers and phonons \cite{Eleftheriou200320, Ivanchenko2004120, Eleftheriou2005142}.

In addition, when the thermalized nonlinear systems are cooling by boundary dissipations, long-lived discrete breathers can sometimes be observed after a long time relaxation \cite{Tsironis19965225, Reigada2002046607, Bikaki19991234, Piazza2003637,Eleftheriou2005142, Ivanchenko2004120, Eleftheriou2005230, Piazza20019803, Piazza2005145502, Juanico2007238104, Reigada2001066608, Reigada2002467, Reigada2003646}. Once the long-lived discrete breathers are formed, the energy relaxation in these nonlinear systems become very slow. This mechanism has been used to explain the anomalous decay of luminescence in certain doped alkali halides \cite{Schulman2002224101, Mihokova2004016610, Mihokova2010183001}. The discrete breathers which are observed experimentally in coupled antiferromagnetic layers\cite{Schwarz1999223, Sato2004486, Wrubel2005264101} can also be considered as being excited in the relaxation processes. Inspired by these findings, boundary dissipation was also used in optic lattices to make the formation of the long-lived discrete breather and consequently leading to the localization of Bose-Einstein condensates \cite{Livi2006060401, Franzosi20071195, Franzosi2011R89, Ng2009073045, Hennig2010053604, Bai2015selective}. It is numerically revealed that the initial average energy per site of the thermalized lattice must higher than a critical energy to excite the long-lived discrete breather in both one dimensional nonlinear lattices \cite{Tsironis19965225, Reigada2002046607} and two dimensional nonlinear lattices \cite{Bikaki19991234, Piazza2003637, Eleftheriou2005142} with boundary dissipation. Therefore, there also exist an threshold of the initial average energy per site for excitation of the long-lived discrete breather which can remain in nonlinear systems after a long time relaxation. The energy threshold $e_{re}^c$ should be also regarded as the activation free energy with respect to the number of the long-lived discrete breathers which can remain after a long time relaxation as $n_{re}\propto \exp(-e_{re}^c/k_B T)$.  This energy threshold of a long-lived discrete breather can also not be estimated by the previous analytical methods.

Because of the absence of the analytical method, the energy thresholds of discrete breathers in thermal equilibrium and relaxation processes can only be numerically estimated. However, the size-dependence of the energy thresholds can not be obtained from these numerical results. They were thus regarded as same as the previous analytical results \cite{Piazza2003637, Eleftheriou2005142}. Therefore, an analytical method is desired to estimate the energy thresholds of discrete breathers in thermal equilibrium as well as in relaxation processes and thus to evaluate the validity of the previous results.

In this work, the common one dimensional nonlinear lattices whose Hamiltonians admit a Taylor series expansion around its equilibrium point are explicitly investigated. The equations of motion are firstly transformed into the normal coordinates based on the system-plus-reservoir models. And then these equations of motion are solved by using the method of averaging. Finally, we analytically estimate the energy thresholds of discrete breathers in thermal equilibrium and relaxation processes. The obtained threshold of the equilibrium average energy per site for discrete breather in thermal equilibrium is $e_0\sim N^{-2}$. This is same as the previous results \cite{FLACH1996223, Flach19971207} for a single discrete breather in one dimensional nonlinear Hamiltonian systems. The threshold of the initial average energy per site for discrete breather in relaxation process is $\sim N^{-1}$. It is in remarkable agreement with the published numerical results known to us \cite{Tsironis19965225, Reigada2002046607} but is different from the previous result ($\sim N^{-2}$) for a single discrete breather in nonlinear Hamiltonian systems.

The rest of the paper is organized as follows. In Sec. \ref{sec2}, the studied models and the analytical method are presented. The obtained energy thresholds of discrete breathers in thermal equilibrium and relaxation processes are presented in Secs. \ref{sec3-1} and \ref{sec3-2} respectively. These analytical results are compared with the published numerical results known to us in Sec. \ref{sec3-3}. The good agreements are obtained. Finally, we draw the conclusions and discuss the potential applications of our results for experimental detection of discrete breathers in thermal equilibrium and relaxation processes in Sec. \ref{sec5}.

\section{Model and method}\label{sec2}
In this work, we do the explicit calculation only for a one dimensional chain. It consists of $N$ oscillators. Whose Hamiltonian is
\begin{equation}\label{hamiltonian}
H_S=\sum_{n=0}^{N-1}\Big[\frac{\dot{u}_n^2}{2}+V(u_n)+W(u_n-u_{n-1})\Big],
\end{equation}
where $u_n$ is the dimensionless displacement of the $n$th oscillator from equilibrium, the dot denotes the time derivative. The potentials
\begin{equation}\label{onsite}
V(z)=\sum_{\mu=2}^\infty \frac{1}{\mu}v_\mu z^\mu
\end{equation}
and
\begin{equation}\label{nearest}
W(z)=\sum_{\mu=2}^\infty \frac{1}{\mu}\phi_\mu z^\mu
\end{equation}
are the on-site potential and the nearest-neighbor coupling potential respectively. The energy of each oscillator can be expressed as
\begin{equation}\label{energy_on_site}
\epsilon_n=\frac{1}{2}\dot{u}_n^2+V(u_n)+\frac{1}{2}[W(u_n-u_{n-1})+W(u_{n+1}-u_{n})].
\end{equation}
The energy relaxation in the chain is critically dependent on the boundary conditions \cite{Piazza20019803, Piazza2003637}. To study the energy threshold of the long-lived discrete breather in the relaxation process, the free-end ($u_{-1}=u_0$, $u_N=u_{N-1}$) boundary condition is imposed as in Refs. [\onlinecite{Tsironis19965225, Reigada2002046607, Bikaki19991234, Piazza2003637,Eleftheriou2005142, Ivanchenko2004120, Eleftheriou2005230, Piazza20019803, Piazza2005145502, Juanico2007238104, Reigada2001066608, Reigada2002467, Reigada2003646}].

The chain is initially thermalized to reach a thermal equilibrium state at time $t=0$ with the ensemble average energy per site being $\langle \epsilon_n\rangle= e_0$. Afterward, i.e., at $t>0$, the chain is connected to the zero temperature reservoirs at its ends by adding the boundary dissipations and consequently the energy relaxation starts. Therefore, we can study the energy threshold of discrete breathers in thermal equilibrium at time $t=0$ or by letting the boundary dissipations equal to zero (i.e., switching off the dissipations). When only one discrete breather remains in the chain after a long time relaxation, the threshold of initial average energy per site for this long-lived discrete breather can be calculated. In the relaxation process, the corresponding equations of motion can be derived based on the system-plus-reservoir model \cite{Weiss1999} as
\begin{eqnarray}\label{langevin}
\ddot{u}_n&=&W^\prime(u_{n+1}-u_n)-W^\prime(u_n-u_{n-1})-V^\prime(u_n)\nonumber\\
&&-\gamma \dot{u}_n(\delta_{n,0}+\delta_{n,N-1}),
\end{eqnarray}
where $\gamma$ is the dissipation constant, $\delta$ denotes the Kronecker delta and a prime denotes the derivative of the function with respect to its argument as
\begin{eqnarray}\label{deriv}
W^\prime(u_{n+1}-u_n)&=&\sum_{\mu=2}^\infty \phi_\mu (u_{n+1}-u_n)^{\mu-1},\nonumber\\
W^\prime(u_n-u_{n-1})&=&\sum_{\mu=2}^\infty \phi_\mu (u_n-u_{n-1})^{\mu-1},\nonumber\\
V^\prime(u_n)&=&\sum_{\mu=2}^\infty  v_\mu u_n^{\mu-1}.
\end{eqnarray}
When $\gamma=0$, Eq. \eqref{langevin} is just the equation of motion for Hamiltonian system.

To solve the equations of motion Eq. \eqref{langevin}, it is convenient to introduce the normal coordinates $Q_k(t)$ according to the canonical transformation \cite{Piazza20019803, Flach2005064102, Flach2006036618}
\begin{equation}\label{normal}
u_n(t)=\sum_{k=1}^{N} A_{n,k}Q_k(t)=\sum_{k=1}^{N}\sqrt{\frac{2}{N}} \cos[\frac{k \pi}{N} (n+\frac{1}{2})]Q_k(t).
\end{equation}
Based on the system-plus-reservoir model, the equations of motion Eq. \eqref{langevin} can be transformed into the normal coordinates space (see Appendix \ref{appsec2} for the details) as
\begin{widetext}
\begin{eqnarray}\label{langevin-norm}
\ddot{Q}_k+\omega_k^2 Q_k&=&-\sum_{\mu=3}^\infty \sum_{n=0}^{N-1} \Bigg\{v_\mu A_{n,k}\Big[\sum_{j=1}^{N}A_{n,j} Q_j\Big]^{\mu-1}+ \phi_\mu (A_{n,k}-A_{n-1,k}) \Big[\sum_{j=1}^{N} (A_{n,j}-A_{n-1,j}) Q_j\Big]^{\mu-1}\Bigg\}\nonumber\\
&&-\gamma A_{0,k}\Big[\sum_{j=1}^{N}A_{0,j}\dot{Q}_j\Big]-\gamma A_{N-1,k}\Big[\sum_{j=1}^{N}A_{N-1,j}\dot{Q}_j\Big]\equiv -F_k(Q),
\end{eqnarray}
\end{widetext}
where
\begin{equation}\label{omega}
  \omega_k=\sqrt{v_2+4\phi_2\sin^2\frac{k \pi}{2N}}
\end{equation}
are the linear normal mode frequencies.

Bearing in mind that we are seeking the lower bound energy of a discrete breather and recalling that the previously obtained critical amplitude of the phonon band edge mode for formation the discrete breather is $\sim 1/N$, \cite{FLACH1996223, Flach19971207} the nonlinear terms in the equations of motion are thus small relative to the linear terms. The lager $N$ is, the smaller the nonlinear terms will be. In addition, we let $\gamma$ be small enough. Therefore, $F_k(Q)$ can be treated as a perturbation term and Eq. \eqref{langevin-norm} can be solved by using the method of averaging \cite{nayfeh1995} (see Appendix \ref{appsec3} for details). Consequently, the corresponding nonlinear oscillation frequencies $\Omega_k$'s can be obtained. When some nonlinear oscillation frequencies $\Omega_k$'s lie outside the phonon spectrum, the corresponding nonlinear oscillation modes are supposed to be discrete breathers according to Refs. [\onlinecite{Flach19971207, Kastner2004104301, Gershgorin2005264302}].

We formally let
\begin{equation}\label{qk}
  {Q}_k(t)={a}_k(t)\cos[\omega_k t+{\beta}_k(t)]={a}_k(t)\cos[{\theta}_k(t)]
\end{equation}
and
\begin{equation}\label{eq11}
\dot{a}_k(t) \cos[{\theta}_k(t)]-a_k(t)\dot{\beta}_k(t) \sin [{\theta}_k(t)]=0.
\end{equation}
It is thus obtained that
\begin{equation}\label{qkdot}
  \dot{{Q}}_k(t)=-\omega_k {a}_k(t)\sin[{\theta}_k(t)].
\end{equation}
By using Eq. \eqref{eq11} as well as substituting $\dot{{Q}}_k(t)$ and $\ddot{{Q}}_k(t)$ into Eq. \eqref{langevin-norm}, it is then obtained that(for the sake of brevity the explicit time dependence is omitted here and the following)
\begin{equation}\label{alpha}
  \dot{{a}}_k=\frac{\sin({\theta}_k)}{\omega_k} F_k({Q})
\end{equation}
and
\begin{equation}\label{beta}
  \dot{{\beta}}_k=\frac{{\cos({\theta}_k)}}{\omega_k {a}_k}F_k({Q}).
\end{equation}
We should mention here that Eqs. \eqref{alpha} and \eqref{beta} are obviously not suitable for $\omega_k=0$. This can take place when $v_2=0$. However, discrete breather can only exist with frequency being higher than the upper band edge frequency when $v_2=0$. Therefore, the energy thresholds of the discrete breathers are determined by the highest frequency mode rather than the mode with the lowest frequency.

Because $F_k({Q})$ is small, $\dot{{a}}_k$ and $\dot{{\beta}}_k$ are small also and thus can be approximated by their time average values \cite{nayfeh1995} (see Appendix \ref{appsec3} for details) as
\begin{eqnarray}
\dot{a}_k &=&\idotsint_0^{2\pi} \frac{\sin\theta_k}{(2\pi)^N\omega_k} F_k(Q) d\theta_1\cdots d\theta_{N}\nonumber\\
&=& -\frac{2\gamma a_k}{N}\cos^2\Big(\frac{k\pi}{2N}\Big), \label{parameter1}
\end{eqnarray}
\begin{widetext}
\begin{eqnarray}
\dot{\beta}_k&=&\idotsint_0^{2\pi}\frac{\cos\theta_k}{(2\pi)^N\omega_k a_k}F_k(Q)d\theta_1\cdots d\theta_{N} = \frac{1}{8N\omega_k}\Bigg\{\sum_{m\neq k}a_m^2\Big[\frac{\phi_4}{\phi_2^2}D_{kkmm}(\omega_k^2-v_2) (\omega_m^2-v_2)\nonumber\\
&&+v_4 C_{kkmm}\Big]+\frac{3}{2}a_k^2\Big[\frac{\phi_4}{\phi_2^2}D_{kkkk}(\omega_k^2-v_2)^2+v_4 C_{kkkk} \Big]\Bigg\},\label{parameter2}
\end{eqnarray}
\end{widetext}
where the coupling coefficients are \cite{BIVINS197365,Luca1995283, Flach2005064102, Flach2006036618}
\begin{eqnarray}\label{coeffc}
  C_{klmp}&=&\Delta'_{k+l+m+p}+\Delta'_{k+l+m-p}+\Delta'_{k+l-m+p}+\Delta'_{k+l-m-p}\nonumber\\
  &&+\Delta'_{k-l+m+p}+\Delta'_{k-l+m-p}+\Delta'_{k-l-m+p}+ \Delta'_{k-l-m-p}
\end{eqnarray}
and
\begin{eqnarray}\label{coeffd}
    D_{klmp}&=&\Delta_{k+l+m+p}+\Delta_{k+l-m-p}+\Delta_{k-l+m-p}+\Delta_{k-l-m+p}\nonumber\\
  &&-\Delta_{k+l+m-p}-\Delta_{k+l-m+p}-\Delta_{k-l+m+p}- \Delta_{k-l-m-p}
\end{eqnarray}
with $\Delta'_r$ and $\Delta_r$ are defined by
\begin{equation} \label{delta1}
\Delta'_r=\left\{ \begin{aligned}
         1 &\qquad\text{for}\quad r=0, \pm 4N \\
         -1&\qquad\text{for}\quad r=\pm 2N \\
         0 & \qquad \text{otherwise},
                  \end{aligned} \right.
\end{equation}
and
\begin{equation} \label{delta2}
\Delta_r=\left\{ \begin{aligned}
         1 &\qquad\text{for}\quad r=0,\pm 2N, \pm 4N \\
         0 & \qquad \text{otherwise}.
                  \end{aligned} \right.
\end{equation}
One can find that
\begin{equation}\label{edge}
C_{NNkk}=C_{kkNN}=D_{NNkk}=D_{kkNN}=0
\end{equation}
for any $k$.
In Eq. \eqref{parameter1}, the nonlinear terms of $Q_j$ contribute nothing to the average values. Only the dissipation terms determine the results. In Eq. \eqref{parameter2}, all the dissipation terms and the nonlinear terms with the odd $\mu$'s contribute nothing to the results. In addition, the sum over $\mu$ of $F_k(Q)$ is truncated at $\mu=4$. Because all the terms with $\mu>4$ are of order $Q_k^4$ or higher and can be neglected when $Q_k$ is small.

Integrating Eq. \eqref{parameter1}, we obtain
\begin{equation}\label{alpha_result}
  a_k(t)=a_k(0)\exp[-\frac{2\gamma t}{N}\cos^2(\frac{k\pi}{2N})] \equiv a_k(0)\exp(-t/\tau_k),
\end{equation}
where $\tau_k$ can be regarded as the relaxation time of the $k$-th mode. As one can expect, when $\gamma=0$, i.e., at equilibrium without dissipations, the relaxation time is $\tau_k=\infty$. Because the initial state is a classical thermal equilibrium state, it is assumed that the equipartition theorem is valid. Furthermore, due to we are seeking the lower bound energy of a discrete breather, the nonlinearity is small. Therefore, the energy of the $k$-th normal mode can be approximated as $\omega_k^2 a_k^2(0)/2\approx e_0$.
Substituting $a_k$ into Eq. \eqref{parameter2}, $\dot{\beta}_k$ can be expressed as
\begin{eqnarray}\label{betak}
&&\dot{\beta}_k= \frac{1}{8N\omega_k}\Bigg\{\sum_{m\neq k}\frac{2e_0}{\omega_m^2}e^{-2t/\tau_m}\Big[v_4 C_{kkmm}\nonumber\\
&&+\frac{\phi_4}{\phi_2^2}D_{kkmm}(\omega_k^2-v_2) (\omega_m^2-v_2)\Big]+\frac{3}{2} \frac{2e_0}{\omega_k^2}e^{-2t/\tau_k}\Big[v_4 C_{kkkk}\nonumber\\
&&+ \frac{\phi_4}{\phi_2^2}D_{kkkk}(\omega_k^2-v_2)^2\Big]\Bigg\},
\end{eqnarray}
where $a_\alpha^2(0)$ is replaced approximately with $2e_0/\omega_\alpha^2$. It is thus obtained that $\dot{\beta}_N=0$ according to Eq. \eqref{edge}.

\section{Results}\label{sec3}
By invoking the general equipartition theorem and considering the dissipations as perturbations, the nonlinear oscillation frequencies (or it can be named as effective frequencies or renormalized frequencies) can be defined as \cite{berne2000dynamic, Lee2009, Lee20133237} (see Appendix \ref{appsec4} for details)
\begin{eqnarray}\label{eff_freq}
\Omega_k=\sqrt{\frac{\langle\dot{Q}_k^2 \rangle}{\langle Q_k^2 \rangle}}
=\sqrt{\frac{\dot{a}_k^2 \langle \cos^2\theta_k \rangle +a_k^2 \dot{\theta}_k^2 \langle \sin^2\theta_k \rangle}{ a_k^2 \langle \cos^2\theta_k \rangle}}\approx \frac{d\theta_k}{dt}=\omega_k+\dot{\beta}_k,
\end{eqnarray}
where $\langle \cdots\rangle$ denotes the ensemble averaging which has been replaced by the time averaging as in Ref. [\onlinecite{Lee20133237}]. Here we perform the averaging over a period $2\pi$ of $\theta_k$ according to the spirit of the method of averaging. According to Eq. \eqref{parameter1}, $\dot{a}_k^2=\frac{4\gamma^2 a_k^2}{N^2}\cos^4\Big(\frac{k\pi}{2N}\Big)\approx 0$ is obtained for large $N$.

The frequency of a generic discrete breather has to lie outside the phonon spectrum. This frequency will approach to the phonon band edge from outside the phonon band with decreasing energy. Therefore, discrete breather is assumed to appear through a tangent bifurcation from a phonon mode in Refs. [\onlinecite{FLACH1996223, Flach19971207, Kastner2004104301, Kastner20041923}]. The tangent bifurcation from the phonon edge mode gives the energy threshold of discrete breather. The necessary condition of the tangent bifurcation is that the frequency of the band edge mode is repelled outside the phonon band with increasing energy \cite{Kastner20041923}. In this work, we assume that discrete breather takes place once the frequency of the phonon mode being repelled outside the phonon band with increasing energy. The critical energy of the repelling of the frequency of the band edge mode determines the energy threshold of discrete breather. It is not known how to prove this assumption. We can only check the validity of it by comparing our results with the reported numerical results at the end of this section.

Because $\dot{\beta}_N=0$, it is obtained that $\Omega_N=\omega_N=\sqrt{v_2+4\phi_2}$. Therefore, we can estimate the energy threshold by letting the nonlinear oscillation frequency of the $(N-1)$-th mode $\Omega_{N-1}$ be higher than the upper band edge frequency $\omega_E=\omega_N=\sqrt{v_2+4\phi_2}$, i.e. $\Omega_{N-1}>\omega_E$. When $v_2=0$, e.g. the Fermi-Pasta-Ulam-$\beta$ (FPU-$\beta$) lattices studied in Refs.~[\onlinecite{Piazza20019803,Piazza2003637, Reigada2001066608, Reigada2002046607, Reigada2002467, Reigada2003646}], the energy threshold can only be estimated by using $\Omega_{N-1}>\omega_E$.

For the $(N-1)$-th mode, Eq.~\eqref{betak} can be expressed as
\begin{widetext}
\begin{eqnarray}\label{beta_edge}
\dot{\beta}_{N-1}&=& \frac{1}{8N\omega_{N-1}}\Bigg\{\sum_{m\neq 1,N-1,N}\frac{2e_0}{\omega_m^2}e^{-2t/\tau_m}\Big[2v_4 +2\frac{\phi_4}{\phi_2^2}(\omega_{N-1}^2-v_2) (\omega_m^2-v_2)\Big]+\frac{9}{2} \frac{2e_0}{\omega_{N-1}^2}e^{-2t/\tau_{N-1}}\Big[v_4 \nonumber\\
&&+ \frac{\phi_4}{\phi_2^2} (\omega_{N-1}^2-v_2)^2\Big]+\frac{2e_0}{\omega_1^2}e^{-2t/\tau_1}\Big[v_4 + 3\frac{\phi_4}{\phi_2^2} (\omega_{N-1}^2-v_2) (\omega_1^2-v_2)\Big]\Bigg\}\nonumber\\
&\approx & \frac{1}{8N\omega_{N-1}}\Bigg\{\sum_{m=1}^{N-1}2\frac{2e_0}{\omega_m^2}e^{-2t/\tau_m}\Big[v_4  +\frac{\phi_4}{\phi_2^2}(\omega_{N-1}^2-v_2) (\omega_m^2-v_2)\Big]\Bigg\}.
\end{eqnarray}
In the summation of the last equality, the frequency of the $(N-1)$-th mode is the highest frequency. Hence $\omega_m\le \omega_{N-1}$ and
\begin{eqnarray}\label{beta_phi4}
\dot{\beta}_{N-1}&>& \frac{1}{8N\omega_{N-1}^3}\Bigg\{\sum_{m=1}^{N-1} 4e_0 \exp\Big[-\frac{2t}{\tau_0} \cos^2(\frac{m\pi}{2N})\Big]\Big[v_4 +\frac{2\phi_4}{\phi_2}(\omega_{N-1}^2-v_2) (1-\cos(\frac{m\pi}{N}))\Big]\Bigg\}\nonumber\\
&=&\frac{e_0}{2\pi \omega_{N-1}^3}\int_0^{\pi}e^{-\frac{t}{\tau_0}}e^{-\frac{t}{\tau_0}\cos q}\Big[v_4+ \frac{2\phi_4}{\phi_2}(\omega_{N-1}^2-v_2) (1-\cos q)\Big] dq\nonumber\\
&=& \frac{e_0}{2\omega_{N-1}^3}e^{-\frac{t}{\tau_0}} I_0(\frac{t}{\tau_0})\Big[v_4+ \frac{2\phi_4}{\phi_2}(\omega_{N-1}^2 -v_2) \Big]+\frac{e_0}{2\omega_{N-1}^3}e^{-\frac{t}{\tau_0}} I_1(\frac{t}{\tau_0})\frac{2\phi_4}{\phi_2}(\omega_{N-1}^2 -v_2),
\end{eqnarray}
\end{widetext}
where $\tau_0=N/2\gamma$, $I_0$ and $I_1$ are the zeroth-order and the first-order modified Bessel functions \cite{NIST:DLMF}. The second equality is obtained by approximating the sum over $m$ with an integral ($\sum_{m=1}^{N-1}\rightarrow \int_0^\pi\frac{N}{\pi}dq$) for large $N$. \cite{Piazza20019803}

\subsection{Results in thermal equilibrium}\label{sec3-1}
We would like to emphasise here again that the chain is initially thermalized to reach a thermal equilibrium state at time $t=0$ with the ensemble average energy per site being $\langle \epsilon_n\rangle= e_0$. Afterward, i.e., at $t>0$, the chain is connected to the zero temperature reservoirs at its ends by adding the boundary dissipations and consequently the energy relaxation starts. Therefore, we can study the energy threshold of discrete breathers in thermal equilibrium at time $t=0$ or by letting the boundary dissipations equal to zero (i.e., $\gamma=0$). When $\gamma=0$, $\tau_0=\infty$ is obtained. And thus $t/\tau_0=0$. Therefore, by using wether $t=0$ or $\gamma=0$, the chain is in the initial thermal equilibrium state when $t/\tau_0=0$.

We can estimate the energy threshold of discrete breathers in thermal equilibrium state according to $\Omega_{N-1}(t/\tau_0=0)\ge \sqrt{v_2+4\phi_2}$. This means $\dot{\beta}_{N-1}(t/\tau_0=0)\ge \sqrt{v_2+4\phi_2}-\omega_{N-1}$ based on Eq. \eqref{eff_freq}. According to Eq. \eqref{beta_phi4} and using $I_0(0)=1$ and $I_1(0)=0$, we simply let
\begin{equation}\label{equilibrium}
\frac{e_0}{2\omega_{N-1}^3}\Big[v_4+ \frac{2\phi_4}{\phi_2}(\omega_{N-1}^2 -v_2) \Big]\ge \sqrt{v_2+4\phi_2}-\omega_{N-1},
\end{equation}
where $\omega_{N-1}=\sqrt{v_2+4\phi_2\sin^2\frac{(N-1)\pi}{2N}}=\sqrt{v_2+4\phi_2\cos^2(\pi/2N)}$. When $N$ is large, $\pi/2N$ is an infinitesimal. It is thus obtained that
\begin{equation}\label{threshold1}
e_0\ge e_{eq}^c=\frac{1}{N^2}\frac{v_2+4\phi_2}{4(v_4+8\phi_4)}4\pi^2\phi_2.
\end{equation}
Where $e_{eq}^c$ is the desired energy threshold (lower bound energy) to excite the discrete breather in the thermal equilibrium state. It should be better regarded as the activation free energy with respect to the number of discrete breathers in thermal equilibrium as $n_{eq}\propto \exp(-e_{eq}^c/k_B T)$. The size-dependence of the energy threshold agrees well with the previous result of energy threshold of a single discrete breather in a Hamiltonian system \cite{FLACH1996223, Flach19971207}. This indicates that the interactions between breathers and phonons do not alter the size-dependence of energy thresholds of discrete breathers. To excite only a single discrete breather (previous results) or a discrete breather among phonons (in thermal equilibrium states) in a Hamiltonian system, the size-dependencies of the energy thresholds are identical.

\subsection{Results in relaxation process}\label{sec3-2}
When $t>0$ and $\gamma>0$, the dissipation is switched on and thus the energy relaxation takes place.
In the relaxation process, as shown in Eq. \eqref{alpha_result}, the lower $k$ is, the relaxation time of the $k$-th mode is shorter. Consequently, it decays faster. To calculate the energy threshold of a long-lived discrete breather which can remain after a long time relaxation, we let that only the $(N-1)$-th mode remains higher than the band edge after a long time relaxation, i.e. $\Omega_{N-1}(t)>\sqrt{v_2+4\phi_2}$ at $t=\tau_{N-1}$. By using Eq. \eqref{beta_phi4} and realizing that $\tau_{N-1}/\tau_0=\sin^{-2}(\frac{\pi}{2N}) \rightarrow \infty$ for large $N$, as Eq. \eqref{equilibrium}, we obtain
\begin{equation}\label{relaxation}
\frac{e_0}{2\omega_{N-1}^3}\frac{1}{\sqrt{2\pi}}\sin\frac{\pi}{2N}\Big[v_4+ \frac{4\phi_4}{\phi_2}(\omega_{N-1}^2 -v_2) \Big]\ge \sqrt{v_2+4\phi_2}-\omega_{N-1},
\end{equation}
where $e^{-z}I_0(z)=e^{-z}I_1(z)=1/\sqrt{2\pi z}$ when $z\rightarrow \infty$ is used. Therefore, it is obtained that
\begin{equation}\label{threshold2}
e_0\ge e_{re}^c=\frac{1}{N}\frac{v_2+4\phi_2}{v_4+16\phi_4}\sqrt{\frac{2}{\pi}}2\pi^2\phi_2.
\end{equation}
Where $e_{re}^c$ is the threshold of the initial average energy per site for the long-lived discrete breather in the relaxation process. It should be also regarded as the activation free energy with respect to the number of the long-lived discrete breathers which can remain after a long time relaxation as $n_{re}\propto \exp(-e_{re}^c/k_B T)$. This result can be understood as follows. After a long time relaxation to $t=\tau_{N-1}$, the initial average energy per site $e_0$ decays to $e_0/\sqrt{2\pi \tau_{N-1}/\tau_0}\sim e_0/N$. (This can also be understood as that the initial energy is equipartition in $N$ normal modes. After a long time relaxation, only the highest normal mode remains. Therefore, the remaining average energy per site is $e_0/N$.) This remaining energy has to be higher than the energy threshold ($\sim 1/N^2$) \cite{FLACH1996223, Flach19971207} of a single discrete breather to excite it \cite{Hajnal2008124101}. Therefore, $e_{re}^c\sim 1/N$ is obtained.

The threshold of the initial average energy per site for the long-lived discrete breather which can remain after a long time relaxation in one dimensional $N$-sites nonlinear chains with boundary dissipation is $\sim 1/N$. This is very different from the energy threshold of a discrete breather in thermal equilibrium as shown in Eq. \eqref{threshold1} and the previous results ($\sim 1/N^2$) \cite{FLACH1996223, Flach19971207} of a single discrete breather in one dimensional Hamiltonian systems.

\subsection{Validity of the results}\label{sec3-3}
The energy threshold of a discrete breather in thermal equilibrium as shown in Eq. \eqref{threshold1} and the energy threshold of a long-lived discrete breather in relaxation process as shown in Eq. \eqref{threshold2} are the central results of this work. The validity of them will be checked in this subsection by comparing them with the reported numerical results. The good agreements are obtained.

It is difficult to numerically identify the discrete breathers in thermal equilibrium \cite{Eleftheriou2005142}. There is thus not a comparable result of the energy threshold of a discrete breather in thermal equilibrium. Therefore, in this subsection, we compare our results of the energy thresholds of long-lived discrete breathers in relaxation processes with the published numerical results. Our analytical results agree well with the published numerical results of one dimensional lattices known to us \cite{Tsironis19965225, Reigada2002046607}.

In Ref. [\onlinecite{Tsironis19965225}] (see figure 2 of it), the parameters of hard $\varphi^4$ chains are $\phi_2=0.1$, $\phi_4=0$, $v_2=v_4=1$ and $N=72$ (wherein $8$ oscillators are contacted to the baths). When the hard $\varphi^4$ chain was thermalized initially to the temperature $T=0.01$, the energy relaxation is exponential as same as it in harmonic chains. However, when the initial temperatures were thermalized to $T=0.1$ and $T=1$, the energy relaxation is distinctly slower than the exponential fashion. The higher the initial temperature is, the slower the energy relaxation is. These numerical results indicate that the threshold of the initial temperature to observe the slow energy relaxation is lying between $0.01$ and $0.1$. According to the equipartition theorem, when the initial average energy per site is low, it equals approximately to the initial thermal energy $T$ (by letting the Boltzmann constant $k_B=1$). Substituting the parameters into Eq. \eqref{threshold2}, the energy threshold is obtained as $e_{re}^c\approx 0.031$. Therefore, the long-lived discrete breathers can remain in the chain and consequently cause the slow energy relaxation merely when the initial temperature is higher than $0.031$. This energy threshold, $0.031$, is indeed lying between $0.01$ and $0.1$.

However, our results agree less well with the numerical results of the soft Morse on-site potential $V(u_n)=\frac{1}{2}[1-\exp(-u_n)]$ in Ref. [\onlinecite{Tsironis19965225}] (see figure 4 of it). The number of sites is also $N=72$. The nearest-neighbor coupling potential is harmonic and thus $\phi_4=0$. By applying Taylor expansion to the soft Morse potential, one can obtain the parameters with $v_2=1$, $v_3=-3/2$ and $v_4=7/6$. The initial temperature is fixed at $T=0.001$. When $\phi_2=0.01$ and $0.05$, the energy relaxation is distinctly slow. By substituting these parameters into Eq. \eqref{threshold2}, it is obtained that $e_{re}^c\approx 0.002$ and $0.011$ for $\phi_2=0.01$ and $0.05$ respectively. Comparing with $T=0.001$, the worse agreements are attributed to the simplicity of the method of averaging used in this work and thus the parameter $v_3$ contributes nothing. To reveal the effect of $v_3$ on the energy threshold, we simply use the equation (3.13) of Ref. [\onlinecite{FLACH1996223}] to estimate the energy threshold by replacing $1/N^2$ with $1/N$ according to the aforementioned discussion. It is obtained that $e_{re}^c\approx 0.001$ and $0.007$ for $\phi_2=0.01$ and $0.05$ respectively. The agreements are distinctly improved. This convince us that the energy threshold of long-lived discrete breathers in a one dimensional $N$-sites nonlinear chain is $\sim 1/N$.

The relaxation of a discrete breather in one dimensional FPU-$\beta$ chains is studied in Ref. [\onlinecite{Reigada2002046607}] by initially creating an ``odd parity'' excitation with amplitudes $u_{n-1}=-A/2$, $u_n=A$, and $u_{n+1}=A/2$ on three successive sites, zero amplitude on the other sites, and zero velocity at each site. Then the total initial energy can be calculated from Eq. \eqref{energy_on_site} as $E_{t}=\sum_{i=n-2}^{n+2} \epsilon_i=\frac{3}{2}\phi_2 A^2+\frac{21}{16}\phi_4 A^4$. Although the initial state of the relaxation is a single discrete breather rather than a thermal equilibrium state. We expect that our result Eq. \eqref{threshold2} is suitable for the relaxation of a single discrete breather because the energy threshold of a discrete breather in thermal equilibrium is same as it of a single discrete breather in a Hamiltonian system. As shown in figure 4 of Ref. [\onlinecite{Reigada2002046607}], the created excitation with $A=0.5$ remains stationary and thus decays very slowly in the chain with $N=31$, but it decays very sharply in the chain with $N=21$. The other parameters are set to $v_2=v_4=0$ and $\phi_2=\phi_4=0.5$. Using these parameters, the average energy of the three successive sites, $E_t/3$, is obtained as $0.076$. Comparing with the energy thresholds calculated from Eq. \eqref{threshold2}, it is higher than the energy threshold ($\approx 0.064$) of the $31$-sites chain but lower than the energy threshold ($\approx 0.094$) of the $21$-sites chain. Therefore, the created discrete breather can be long-lived in the $31$-sites chain but decays sharply in the $21$-sites chain. The remarkable agreements with our results indicate that Eq. \eqref{threshold2} is also suitable for the relaxation of a single discrete breather.

\section{Conclusion and discussion} \label{sec5}
In conclusion, for one dimensional lattices, according to the system-plus-reservoir models, by transforming the equations of motion into normal coordinates and sequentially solving them by using the method of averaging, we have obtained the energy thresholds of discrete breathers in thermal equilibrium and the energy thresholds of long-lived discrete breathers which can remain in nonlinear systems after a long time relaxation. The energy thresholds ($\sim N^{-2}$) of discrete breathers in thermal equilibrium are same as the previous results for the energy thresholds of single discrete breathers in one dimensional Hamiltonian systems. The thresholds of the initial average energy per site ($\sim N^{-1}$) for long-lived discrete breathers in relaxation processes are different from the previous results for the energy thresholds of single discrete breathers in one dimensional Hamiltonian systems. However, the results agree well with the published numerical results of one dimensional lattices known to us.

The energy thresholds of discrete breathers in thermal equilibrium are size dependent. The size-dependence can be used to experimentally distinguish the discrete breathers from the localization induced by disorder. In addition, once a long-lived discrete breather can remain after a long time relaxation, a finite portion of energy is localized by it and consequently the energy relaxation is slow. Accordingly, the obtained energy thresholds of the long-lived discrete breathers in relaxation processes are also the energy thresholds for the slow energy relaxation in nonlinear systems. Therefore, the energy thresholds of the long-lived discrete breathers in relaxation processes are expected to be a criterion for the formation of discrete breathers in nonlinear systems by detecting the size-dependence of the energy thresholds of slow energy relaxation.

\begin{acknowledgments}
We thank the referees for their constructive comments. Z.-J.D. is supported by the National Natural Science Foundation of China (Grant Nos. 11274288 and 11574289).
\end{acknowledgments}
\appendix

\section{Equations of motion in normal coordinates space\label{appsec2}}
According to the system plus reservoir model \cite{Weiss1999}, the Langevin equations of motion Eq. \eqref{langevin} can be derived from the Hamiltonian (For the sake of simplicity, the derivation is performed only for a system coupling to one reservoir. The results can be extended straightforwardly to a multi-reservoir system.)
\begin{equation}\label{eq1}
H=H_S+\sum_\alpha \Big[\frac{p_\alpha^2}{2m_\alpha}+\frac{m_\alpha \nu_\alpha^2}{2}(q_\alpha-\frac{c_\alpha}{m_\alpha \nu_\alpha^2}u_c)^2\Big],
\end{equation}
where $H_S$ is the Hamiltonian of the one dimensional chain which is expressed as Eq. \eqref{hamiltonian}. The reservoir (heat bath) consists of a set of harmonic oscillators, whose coordinates and the corresponding momenta are $q_\alpha$ and $p_\alpha$ respectively. The reservoir is coupled to the chain at the $c$-th oscillator whose coordinate is $u_c$.

This Hamiltonian can be transformed into normal coordinates $\{Q_k\}$ by using the canonical transformation \cite{Piazza20019803, Flach2005064102, Flach2006036618}
\begin{equation}\label{eq14}
u_n(t)=\sum_{k=1}^{N} A_{n,k}Q_k(t),
\end{equation}
where
\begin{widetext}
\begin{equation}\label{eq15}
A_{n,k}=\sqrt{\frac{2}{N}}\cos[q_k(n+\frac{1}{2})]\qquad q_k=\frac{k\pi}{N} \qquad k=1,2,\cdots,N
\end{equation}
for the free-end chain, and
\begin{equation}\label{eq16}
A_{n,k}=\sqrt{\frac{2}{N}}\sin[q_k(n+1)]\qquad q_k=\frac{k+1}{N+1}\pi \qquad k=1,2,\cdots,N
\end{equation}
for the fixed-end chain. The Hamiltonian Eq. \eqref{eq1} is thus transformed as
\begin{eqnarray}\label{eq17}
H&=&\sum_{n=0}^{N-1}\Bigg\{\frac{1}{2}\Big[\sum_{k=1}^{N}A_{n,k}\dot{Q}_k\Big]^2+\sum_{\mu=2}^\infty \frac{1}{\mu}v_\mu \Big[\sum_{k=1}^{N}A_{n,k} Q_k\Big]^\mu +\sum_{\mu=2}^\infty \frac{1}{\mu}\phi_\mu\Big[\sum_{k=1}^{N} (A_{n,k}-A_{n-1,k}) Q_k\Big]^\mu\Bigg\}\nonumber\\
&& +\sum_\alpha \Bigg\{\frac{p_\alpha^2}{2m_\alpha}+\frac{m_\alpha \nu_\alpha^2}{2}\Big[q_\alpha-\frac{c_\alpha}{m_\alpha \nu_\alpha^2} \Big(\sum_{k=1}^{N}A_{c,k} Q_k\Big)\Big]^2\Bigg\}\nonumber\\
&=&\sum_{k=1}^{N}\Big[\frac{\dot{Q}_k^2}{2}+\frac{1}{2}\omega_k^2 Q_k^2\Big] +\sum_{\mu=3}^\infty \sum_{n=0}^{N-1} \Bigg\{\frac{1}{\mu}v_\mu \Big[\sum_{k=1}^{N}A_{n,k} Q_k\Big]^\mu+ \frac{1}{\mu}\phi_\mu\Big[\sum_{k=1}^{N} (A_{n,k}-A_{n-1,k}) Q_k\Big]^\mu\Bigg\}\nonumber\\
&& +\sum_\alpha \Bigg\{\frac{p_\alpha^2}{2m_\alpha}+\frac{m_\alpha \nu_\alpha^2}{2}\Big[q_\alpha-\frac{c_\alpha}{m_\alpha \nu_\alpha^2} \Big(\sum_{k=1}^{N}A_{c,k} Q_k\Big)\Big]^2\Bigg\},
\end{eqnarray}
\end{widetext}
where $\omega_k=\sqrt{v_2+4\phi_2\sin^2(q_k/2)}$ are the linear normal mode frequencies.

Therefore, the equations of motion in normal coordinates space are
\begin{eqnarray}
&&\ddot{Q}_k=-\omega_k^2 Q_k-\sum_{\mu=3}^\infty \sum_{n=0}^{N-1} \Bigg\{v_\mu A_{n,k}\Big[\sum_{j=1}^{N}A_{n,j} Q_j\Big]^{\mu-1}\nonumber\\
&&+ \phi_\mu (A_{n,k}-A_{n-1,k}) \Big[\sum_{j=1}^{N} (A_{n,j}-A_{n-1,j}) Q_j\Big]^{\mu-1}\Bigg\}\nonumber\\
&&+A_{c,k} \sum_\alpha c_\alpha \Big[q_\alpha-\frac{c_\alpha}{m_\alpha \nu_\alpha^2}\Big(\sum_{j=1}^{N}A_{c,j}Q_j\Big)\Big],\label{eq18a}\\
&&m_\alpha \ddot{q}_\alpha = -m_\alpha \nu_\alpha^2 q_\alpha +c_\alpha \Big(\sum_{j=1}^{N}A_{c,j} Q_j\Big).\label{eq18b}
\end{eqnarray}
According to Ref. [\onlinecite{Weiss1999}], it can be obtained that
\begin{eqnarray}\label{eq19}
&&\ddot{Q}_k=-\omega_k^2 Q_k-\sum_{\mu=3}^\infty \sum_{n=0}^{N-1} \Bigg\{v_\mu A_{n,k}\Big[\sum_{j=1}^{N}A_{n,j} Q_j\Big]^{\mu-1}\nonumber\\
&&+ \phi_\mu (A_{n,k}-A_{n-1,k}) \Big[\sum_{j=1}^{N} (A_{n,j}-A_{n-1,j}) Q_j\Big]^{\mu-1}\Bigg\}\nonumber\\
&&+A_{c,k}\Bigg\{-\int_{t_0}^t \gamma(t-s) \Big[\sum_{j=1}^{N}A_{c,j}\dot{Q}_j(s)\Big] ds +\xi(t)\Bigg\}.
\end{eqnarray}
For strict ohmic dissipation, the equations of motion are
\begin{eqnarray}\label{eq20}
&&\ddot{Q}_k+\omega_k^2 Q_k=-\sum_{\mu=3}^\infty \sum_{n=0}^{N-1} \Bigg\{v_\mu A_{n,k}\Big[\sum_{j=1}^{N}A_{n,j} Q_j\Big]^{\mu-1}\nonumber\\
&&+ \phi_\mu (A_{n,k}-A_{n-1,k}) \Big[\sum_{j=1}^{N} (A_{n,j}-A_{n-1,j}) Q_j\Big]^{\mu-1}\Bigg\}\nonumber\\
&&-\gamma A_{c,k}\Big[\sum_{j=1}^{N}A_{c,j}\dot{Q}_j\Big]\equiv -F_k(Q)
\end{eqnarray}
for zero temperature reservoir with $T=0$
\section{The method of averaging\label{appsec3}}
When $F_k(Q)$ is small enough, Eq. \eqref{eq20} can be solved perturbatively by using the method of averaging \cite{nayfeh1995}. We formally let
\begin{eqnarray}
Q_k(t)=a_k(t)\cos[\omega_k t+\beta_k(t)]=a_k(t)\cos[\theta_k(t)]. \label{eq21a}
\end{eqnarray}
and
\begin{equation}\label{eq21c}
\dot{a}_k(t) \cos[{\theta}_k(t)]-a_k(t)\dot{\beta}_k(t) \sin [{\theta}_k(t)]=0.
\end{equation}
It is thus obtained that
\begin{eqnarray}
\dot{Q}_k(t)=-\omega_k a_k(t)\sin[\theta_k(t)]. \label{eq21b}
\end{eqnarray}
By using Eq. \eqref{eq21c} as well as substituting $\dot{Q}_k(t)$ and $\ddot{Q}_k(t)$ into Eq. \eqref{eq20},  it is then obtained that
\begin{eqnarray}
\dot{a}_k&=&\frac{\sin\theta_k}{\omega_k} F_k(Q), \label{eq22a}\\
\dot{\beta}_k&=&\frac{\cos\theta_k}{\omega_k a_k} F_k(Q).\label{eq22b}
\end{eqnarray}
For small $F_k(Q)$, $\dot{a}$ and $\dot{\beta}$ are thus small. Therefore, $\dot{a}_k$ and $\dot{\beta}_k$ can be approximated by their time average values \cite{nayfeh1995}.

According to the spirit of the method of averaging, $\dot{a}_k$ is replaced with its average value
\begin{widetext}
\begin{eqnarray}\label{eq23}
\dot{a}_k&=&\idotsint_0^{2\pi} \frac{\sin\theta_k}{(2\pi)^N\omega_k} F_k(Q) d\theta_1\cdots d\theta_{N} \nonumber\\ &=&\idotsint_0^{2\pi} \frac{\sin\theta_k}{(2\pi)^N\omega_k} \gamma A_{c,k}\Big[-\sum_{j=0}^{N-1}A_{c,j}\omega_j a_j\sin\theta_j \Big] d\theta_1\cdots d\theta_{N}\nonumber\\
&=&-\int_0^{2\pi} \frac{\gamma A_{c,k}^2 a_k}{2\pi}\sin^2\theta_k d\theta_k=-\frac{\gamma}{2}A_{c,k}^2 a_k.
\end{eqnarray}
\end{widetext}
It can be obtained that
\begin{equation}\label{eq24}
a_k(t)=a_k(0)\exp(-\frac{\gamma}{2}A_{c,k}^2 t).
\end{equation}
This result can be straightforwardly extended to a multi-reservoir system with
\begin{equation}\label{eq25}
a_k(t)=a_k(0)\exp(-\frac{\gamma}{2}\sum_c A_{c,k}^2 t).
\end{equation}
This is consistent with the result of Ref. [\onlinecite{Piazza20019803}].

$\dot{\beta}_k$ is also replaced with its average value
\begin{eqnarray}\label{eq26}
\dot{\beta}_k&=&\idotsint_0^{2\pi}\frac{\cos\theta_k}{(2\pi)^N\omega_k a_k}F_k(Q)d\theta_1\cdots d\theta_{N}.
\end{eqnarray}
One can obtain that the terms of the sum with odd $\mu$'s in Eq. \eqref{eq20} contribute nothing to $\dot{\beta}$ after doing the averaging. In addition, to calculate the averaging Eq. \eqref{eq26}, the sum over $\mu$ in $F_k(Q)$ is truncated at $\mu=4$. All the terms with $\mu>4$ are of order $Q_k^4$ or higher and can be neglected when $Q_k$ is small. Therefore, Eq. \eqref{eq26} can be expressed as
\begin{widetext}
\begin{eqnarray}\label{eq27}
\dot{\beta}_k&=&\idotsint_0^{2\pi}d\theta_1\cdots d\theta_{N}\frac{\cos\theta_k}{(2\pi)^N\omega_k a_k} \sum_{j=1}^{N}\sum_{l=1}^{N}\sum_{m=1}^{N} a_ja_la_m\cos\theta_j \cos\theta_l \cos\theta_m\nonumber\\
&&\times\Bigg\{v_4 \sum_{n=0}^{N-1}A_{n,k}A_{n,j}A_{n,l}A_{n,m}\nonumber\\
&&+\phi_4 \sum_{n=0}^{N-1} (A_{n,k}-A_{n-1,k}) (A_{n,j}-A_{n-1,j}) (A_{n,l}-A_{n-1,l}) (A_{n,m}-A_{n-1,m})\Bigg\}.
\end{eqnarray}
For free-end chain, one can obtain \cite{BIVINS197365,Luca1995283, Flach2005064102, Flach2006036618}
\begin{eqnarray}\label{eq28}
\sum_{n=0}^{N-1}A_{n,k}A_{n,j}A_{n,l}A_{n,m}=\frac{1}{2N}(\Delta'_{k+j+l+m}+\Delta'_{k+j+l-m}+\Delta'_{k+j-l+m}
+\Delta'_{k+j-l-m}\nonumber\\
  +\Delta'_{k-j+l+m}+\Delta'_{k-j+l-m}+\Delta'_{k-j-l+m}+ \Delta'_{k-j-l-m})\equiv \frac{1}{2N} C_{kjlm}
\end{eqnarray}
and
\begin{eqnarray}\label{eq29}
&&\sum_{n=0}^{N-1}(A_{n,k}-A_{n-1,k}) (A_{n,j}-A_{n-1,j}) (A_{n,l}-A_{n-1,l}) (A_{n,m}-A_{n-1,m})\nonumber\\
&&=\frac{16}{2N}\sin\frac{q_k}{2}\sin\frac{q_j}{2}\sin\frac{q_l}{2}\sin\frac{q_m}{2}
\times\Bigg[\Delta_{k+j+l+m}+\Delta_{k+j-l-m}+\Delta_{k-j+l-m}+\Delta_{k-j-l+m}\nonumber\\
  &&-\Delta_{k+j+l-m}-\Delta_{k+j-l+m}-\Delta_{k-j+l+m}- \Delta_{k-j-l-m}\Bigg]\nonumber\\
&&\equiv \frac{D_{kjlm}}{2N\phi_2^2} \sqrt{\omega_k^2-v_2}\sqrt{\omega_j^2-v_2} \sqrt{\omega_l^2-v_2}\sqrt{\omega_m^2-v_2},
\end{eqnarray}
where $\Delta'_r$ and $\Delta_r$ are defined by
\begin{equation} \label{eq28-1}
\Delta'_r=\left\{ \begin{aligned}
         1 &\qquad\text{for}\quad r=0, \pm 4N \\
         -1&\qquad\text{for}\quad r=\pm 2N \\
         0 & \qquad \text{otherwise},
                  \end{aligned} \right.
\end{equation}
and
\begin{equation} \label{eq29-1}
\Delta_r=\left\{ \begin{aligned}
         1 &\qquad\text{for}\quad r=0,\pm 2N, \pm 4N \\
         0 & \qquad \text{otherwise}.
                  \end{aligned} \right.
\end{equation}
Substituting them into Eq. \eqref{eq27}, it is obtained that
\begin{eqnarray}\label{eq30}
\dot{\beta}_k&=&\idotsint_0^{2\pi}d\theta_1\cdots d\theta_{N}\frac{\cos\theta_k}{(2\pi)^N\omega_k a_k} \sum_{j=1}^{N}\sum_{l=1}^{N}\sum_{m=1}^{N} a_ja_la_m\cos\theta_j \cos\theta_l \cos\theta_m\Bigg\{ \frac{v_4}{2N} C_{kjlm}\nonumber\\
&&+\frac{\phi_4 D_{kjlm}}{2N\phi_2^2} \sqrt{\omega_k^2-v_2}\sqrt{\omega_j^2-v_2} \sqrt{\omega_l^2-v_2}\sqrt{\omega_m^2-v_2}\Bigg\}\nonumber\\
&=&\frac{1}{2N\omega_k}\Bigg\{\sum_{m\neq k}a_m^2\int_0^{2\pi}\frac{\cos^2\theta_k}{2\pi}d\theta_k \int_0^{2\pi}\frac{\cos^2\theta_m}{2\pi}d\theta_m\Big[v_4C_{kkmm}+\frac{\phi_4}{\phi_2^2}D_{kkmm} (\omega_k^2-v_2)(\omega_m^2-v_2)\Big]\nonumber\\
&&+a_k^2 \int_0^{2\pi}\frac{\cos^4\theta_k}{2\pi}d\theta_k \Big[v_4C_{kkkk}+\frac{\phi_4}{\phi_2^2}D_{kkkk} (\omega_k^2-v_2)^2\Big]\Bigg\}\nonumber\\
&=&\frac{1}{8N\omega_k}\Bigg\{\sum_{m\neq k}a_m^2\Big[v_4C_{kkmm}+\frac{\phi_4}{\phi_2^2}D_{kkmm} (\omega_k^2-v_2)(\omega_m^2-v_2)\Big]\nonumber\\
&&+\frac{3}{2}a_k^2 \Big[v_4C_{kkkk}+\frac{\phi_4}{\phi_2^2}D_{kkkk} (\omega_k^2-v_2)^2\Big]\Bigg\}.
\end{eqnarray}
\end{widetext}

\section{Nonlinear oscillation frequency}\label{appsec4}
The nonlinear oscillation frequency Eq. \eqref{eff_freq} is obtained by using the Zwanzig-Mori projection formalism. \cite{berne2000dynamic, Lee2009} In an Hamiltonian system, the time evolution of the dynamical variable $A(t)$ is determined by the Hermitian Liouvillian $\mathcal{L}$ as
\begin{equation}\label{appceq1}
\frac{d A(t)}{dt}=\{A(t),H\}=i\mathcal{L} A(t)
\end{equation}
where $\{A(t),H\}$ is the Poisson bracket of $A(t)$ with the Hamiltonian $H$. An operator which projects an arbitrary vector $B$ onto the subspace spanned by $A$ is
\begin{equation}\label{appceq2}
\mathcal{P}B=(B,A^\dag)(A,A^\dag)^{-1}A
\end{equation}
where $A^\dag$ is the Hermitian conjugate of $A$. The scalar product of two variables $B$ and $A$ is defined as $(B,A^\dag)\equiv \langle BA^\dag \rangle$ where $\langle \cdots \rangle$ denotes with the statistical average over the Gibbs measure induced by the given Hamiltonian. Therefore, Eq. \eqref{appceq1} can be reexpressed as
\begin{eqnarray}\label{appceq3}
\frac{d A(t)}{dt}&=&i\mathcal{L} e^{i\mathcal{L}t} A=e^{i\mathcal{L}t} i\mathcal{L} A \nonumber\\
&=&e^{i\mathcal{L}t}\mathcal{P} i\mathcal{L} A \nonumber+e^{i\mathcal{L}t}(\mathcal{I}-\mathcal{P}) i\mathcal{L} A \nonumber\\
&=&-\Omega^2 A(t)-\int_0^t ds K(t-s)A(s) +F(t)
\end{eqnarray}
where $\mathcal{I}$ is the identity operator, $\Omega^2=-(i\mathcal{L}A,A^\dag)(A,A^\dag)^{-1}$, $F(t)=e^{i(\mathcal{I}-\mathcal{P})\mathcal{L}t}(\mathcal{I}-\mathcal{P})i\mathcal{L}A$ and $K(t-s)=(F(t),F^\dag(s))(A,A^\dag)^{-1}$.

For our system with the Hamiltonian Eq. \eqref{eq17}, the application of Eq. \eqref{appceq3} with the column matrix $A=(Q_k,\dot{Q}_k)^T$ leads to
\begin{equation}\label{appceq4}
\ddot{Q}_k(t)=-\Omega_k^2 Q_k(t)-\int_0^t ds K_k(t-s)\dot{Q}_k(s) +F_k(t)
\end{equation}
where $F_k(t)=e^{i(\mathcal{I}-\mathcal{P})\mathcal{L}t}(\mathcal{I}-\mathcal{P})i\mathcal{L}\dot{Q}_k$, $K_k(t-s)=(F(t),F^*(s))/\langle \dot{Q}_k^2\rangle$ and
\begin{equation}\label{appceq5}
\Omega_k^2=\frac{\langle \frac{\partial H}{\partial Q_k} Q_k \rangle}{\langle Q_k^2\rangle}.
\end{equation}
By using the general equipartition law \cite{Tolman1918}:
\begin{equation}\label{appceq6}
\langle \frac{\partial H}{\partial Q_k} Q_k \rangle=\langle \frac{\partial H}{\partial \dot{Q}_k} \dot{Q}_k \rangle=\langle \dot{Q}_k^2 \rangle,
\end{equation}
Eq. \eqref{eff_freq} can be obtained.


%

\end{document}